\documentclass[aps,prl,showpacs,twocolumn,superscriptaddress, floatfix, tightenlines, amsmath, amssymb]{revtex4-1}
\usepackage[pdftex]{graphicx} 

\graphicspath{{figures/}}
\usepackage{natbib}
\usepackage{booktabs}
\usepackage{xcolor}
\usepackage{url}
\usepackage{amsmath}
\usepackage{float}
\usepackage[colorlinks=true, citecolor=blue, linkcolor=blue, urlcolor=blue]{hyperref}
\usepackage{multirow}
\usepackage{array}
\usepackage{xfrac}
\usepackage{xparse}
\usepackage{etoolbox}
\usepackage{comment}
\usepackage{physics}

\NewDocumentCommand{\Pyrochlore}{O{} O{} O{} O{} O{} O{}}{%
  \textit{\IfBlankTF{#1}{A}{#1}}%
  \SupSub{\IfBlankTF{#2}{}{#2}}{2}%
  \textit{\IfBlankTF{#3}{B}{#3}}%
  \SupSub{\IfBlankTF{#4}{}{#4}}{2}%
  \textit{\IfBlankTF{#5}{X}{#5}}%
  \Sub{\IfBlankTF{#6}{7}{#6}}%
}

\newcommand{\NCNF}{NaCaNi$_2$F$_7$} 
\newcommand{\eloss}{$E_{\mathrm{loss}}$}
\newcommand{\elossR}[2]{\eloss{} = #1 $\rightarrow$ #2\;eV}
\newcommand{\Ei}{E$_i$}
\newcommand{\Sup}[1]{$^{#1}$}
\newcommand{\Sub}[1]{$_{#1}$}
\newcommand{\SupSub}[2]{$^{#1}_{#2}$}
\newcommand{\FrmtSym}[3]{\ensuremath{{}^{#1}#2_{#3}}}

\newcommand{\Pot}[2]{\ensuremath{\hat{\mathcal{V}}_{#1}^{#2}}}
\newcommand{\MuEff}{\ensuremath{\mu_{\rm{eff}}}}

\newcommand{\FigMap}[1]{Figure{#1}~\ref{fig:Overview_MapL23}}

\newcommand{\FigTrigonal}[1]{Figure{#1}~\ref{fig:LFM_Trigonal}}
\newcommand{\FigKedge}[1]{Figure{#1}~\ref{fig:XAS_Kedge}}
\newcommand{\FigDisorder}[1]{Figure{#1}~\ref{fig:Disorder}}

\providecommand{\bibnamefont}[1]{#1}
\providecommand{\bibfnamefont}[1]{#1}
\providecommand{\citenamefont}[1]{#1}
\providecommand{\bibinfo}[2]{#2}
\providecommand{\bibfield}[2]{#2}
\providecommand{\BibitemOpen}{}
\providecommand{\BibitemShut}[1]{}

\providecommand{\Eprint}[2]{\href{#1}{#2}}

\begin{document}
\title{Stability of the Local Ni\Sup{2+} Electronic Structure to \textit{A}-site Disorder in the Pyrochlore Antiferromagnet \NCNF{} }
\author{M. F. DiScala}
\affiliation{Department of Physics, Brown University, Providence, Rhode Island 02912, United States}
\author{A. de la Torre}
\affiliation{Department of Physics, Brown University, Providence, Rhode Island 02912, United States}
\affiliation{Department of Physics, Northeastern University, Boston, Massachusetts 02115, USA}
\affiliation{Quantum Materials and Sensing Institute, Northeastern University, Burlington, Massachusetts 01803, USA}
\author{J. W. Krizan}
\affiliation{Department of Chemistry, Princeton University, Princeton, New Jersey 08544, USA}
\author{J. Wouters}
\affiliation{Department of Physics, Brown University, Providence, Rhode Island 02912, United States}
\author{V. Bisogni}
\affiliation{National Synchrotron Light Source II, Brookhaven National Laboratory, Upton, New York 11973, United States}
\author{J. Pelliciari}
\affiliation{National Synchrotron Light Source II, Brookhaven National Laboratory, Upton, New York 11973, United States}
\author{R. J. Cava}
\affiliation{Department of Chemistry, Princeton University, Princeton, New Jersey 08544, USA}
\author{K. W. Plumb}
\affiliation{Department of Physics, Brown University, Providence, Rhode Island 02912, United States}
\date{\today}

\begin{abstract}
 \NCNF{} is a unique example of spin-1 Heisenberg antiferromagnet on the pyrochlore lattice, but the presence of Na\Sup{1+}/Ca\Sup{2+} \textit{A}-site disorder complicates the local electronic and magnetic environment of the Ni\Sup{2+} \textit{B}-site. We utilize resonant inelastic X-ray scattering (RIXS) to study the influence of \textit{A}-site disorder on the \textit{B}-site electronic structure of \NCNF{}. Ni L-edge RIXS measurements reveal a Ni\Sup{2+} electronic structure in nearly ideal octahedral coordination, with only a small trigonal compression ($\delta$ = -200\;meV) required to capture all spectral features. Within the $D_{3d}$ symmetry of the Ni local environment, we extract an anisotropic $g$-factor of $g_{\parallel} = 2.26$ and $g_{\perp} = 2.27$, and a corresponding paramagnetic moment of $\mu_{\rm{eff}}=3.2\;\mu_B$. To simulate disorder, RIXS spectra were calculated with realistic distributions of crystal field parameters; however, these spectra are invariant relative to a disorder-free model, demonstrating the robustness of the Ni\Sup{2+} electronic environment to the \textit{A}-site disorder, within the resolution of our measurement.  
\end{abstract}
\maketitle

\section{Introduction}

In the pyrochlore lattice, the frustration of antiferromagnetically coupled nearest neighbor spins on corner sharing tetrahedra restricts the development of trivial magnetic ground states \cite{ramirezStronglyGeometricallyFrustrated1994, greedanGeometricallyFrustratedMagnetic2001}. The family of rare-earth transition metal oxides, with formula \Pyrochlore[][3+][][4+][O], has long been a representative of the unique magnetic ground states that emerge on the pyrochlore lattice; these include spin ice, spin glass, and spin liquid phases \cite{machidaTimereversalSymmetryBreaking2010, machidaUnconventionalAnomalousHall2007, nakaiCoreexcitonAbsorptionAbsorption1988, greedanFrustratedRareEarth2006}. However, exploration of the magnetic phases is hindered by the low energy scales of rare-earth magnetism, which requires ultra low temperatures to access the ground state \cite{sibilleCandidateQuantumSpin2015,gaoExperimentalSignaturesThreedimensional2019}.
In contrast, since superexchange interactions in transition metal systems occur at much larger energy scales, integrating a low-spin transition metal into the pyrochlore lattice would enable the study of quantum magnetic phases over a substantially larger temperature range \cite{krizanNaCaCo2F7SinglecrystalHightemperature2014,krizanNaCaNFrustratedHightemperature2015, sandersNaSrMn2F7NaCaFe2F7NaSrFe2F72016, rossSingleionPropertiesSeff2017, gaoSpin1PyrochloreAntiferromagnets2020, kanckoGlassyDisorderedGround2025}.

Recently, a new family of pyrochlore fluorides have been synthesized which meet charge balance by randomly distributing monovalent and divalent cations on the \textit{A}-site \cite{oliveiraCrystalStructureVibrational2004, krizanNaCaCo2F7SinglecrystalHightemperature2014, krizanNaCaNFrustratedHightemperature2015, sandersNaSrMn2F7NaCaFe2F7NaSrFe2F72016, rossSingleionPropertiesSeff2017, gaoSpin1PyrochloreAntiferromagnets2020, kanckoGlassyDisorderedGround2025}. These pyrochlore fluorides accommodate low valence, low spin state transition metals on the \textit{B}-site and have the formula (\textit{A}\SupSub{1+}{\sfrac{1}{2}}\textit{A}$^{\prime}$\SupSub{2+}{\sfrac{1}{2}})\Sub{2}\textit{B}\SupSub{2+}{2}\textit{F}\Sub{7} ($\gamma$-pyrochlores). Unlike the $\beta$-pyrochlore fluorides, which achieve charge balance via a disordered, mixed valence metal \textit{B}-site, the \textit{A}-site disorder in the $\gamma$-pyrochlores results in a \textit{B}-site octahedral environment closer to the ideal and preserves geometric frustration \cite{oliveiraCrystalStructureVibrational2004, krizanNaCaCo2F7SinglecrystalHightemperature2014, krizanNaCaNFrustratedHightemperature2015,  masachchiCrystalGrowthMagnetism2023, kanckoGlassyDisorderedGround2025}. Together, these attributes make the $\gamma$-pyrochlores an ideal platform to explore the unique magnetic phases that can occur in the low spin limit of the pyrochlore lattice \cite{krizanNaCaCo2F7SinglecrystalHightemperature2014, krizanNaCaNFrustratedHightemperature2015, sandersNaSrMn2F7NaCaFe2F7NaSrFe2F72016, rossSingleionPropertiesSeff2017, gaoSpin1PyrochloreAntiferromagnets2020, kanckoGlassyDisorderedGround2025}.

Among these newly synthesized $\gamma$-pyrochlores, \NCNF{} is a unique example of a spin-1 nearest-neighbor Heisenberg antiferromagnet on the pyrochlore lattice \cite{krizanNaCaNFrustratedHightemperature2015, plumbContinuumQuantumFluctuations2019, zhangDynamicalStructureFactor2019}. Inelastic neutron scattering and magnetic susceptibility measurements reveal strong magnetic frustration and signatures of a quantum spin-liquid ground state.  \cite{silversteinLiquidlikeCorrelationsSinglecrystalline2014,krizanNaCaNFrustratedHightemperature2015,caiMSRStudySpin2018, plumbContinuumQuantumFluctuations2019}. The   \textit{A}-site disorder is believed to result in exchange disorder in the magnetic Hamiltonian and ultimately precipitates a spin glass phase below a freezing temperature of T$_{\rm f}\approx 3.6$~K. However, the microscopic impact of \textit{A}-site disorder on the local electronic environment of Ni\Sup{2+} remains largely unresolved. Powder diffraction measurements reveal an $\sim$\;7.5\% deviation of the F-Ni-F bond angles ($83.2^{\circ}$, $96.8^{\circ}$) from an ideal octahedral environment, consistent with a slight trigonal compression \cite{krizanNaCaNFrustratedHightemperature2015}. This distortion may contribute to the anomalously large reported moment ($\mu_{\rm{eff}}$\;=\;$3.7(1)\,\mu_B$) and could similarly be tied to the \textit{A}-site environment, as it has been suggested in rare-earth pyrochlore iridates  \cite{krizanNaCaNFrustratedHightemperature2015, witczak-krempaCorrelatedQuantumPhenomena2014, clancyXrayScatteringStudy2016, krajewskaAlmostPureJeff12020, antonovPyrochloreIridatesElectronic2020}. Detailed electronic structure measurements are therefore necessary to clarify the relationship between \textit{A}-site disorder, the observed trigonal distortion, and their combined influence on the Ni\Sup{2+} electronic and magnetic behavior.  

\begin{figure*}[ht!]
  \begin{center}
    \includegraphics[]{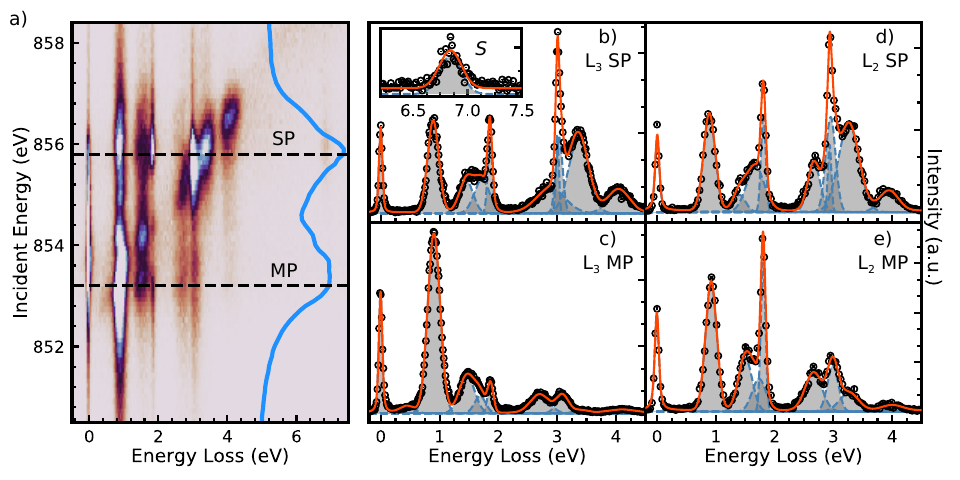}
  \caption{(a) \NCNF{} RIXS map across the Ni L$_3$ as a function of incident energy $E_i$ and energy loss \eloss{}. The solid blue line shows the integrated RIXS intensity (PFY) as a function of $E_i$. 
  (b,\,c) Representative L$_3$ RIXS spectra at \Ei{} = 855.8\;eV and 853.2\;eV, designated SP and MP respectively, and fit to data (red) with Gaussians and Voigt elastic line (dashed blue and filled gray). Inset in (b) shows \eloss{} $\approx$ 7\;eV excitation in SP, denoted peak \textit{S}. (d,\,e) Representative L$_2$ RIXS spectra at $E_i$ = 871.8\;eV and 870.6\;eV}
  \label{fig:Overview_MapL23}
  \end{center}
\end{figure*}

In this work, we use resonant inelastic X-ray scattering (RIXS) to probe the local electronic structure of Ni\Sup{2+} in \NCNF{}. The Ni L\Sub{3}-edge spectra are well described by localized single-ion excitations of Ni\Sup{2+} in octahedral symmetry with a small trigonal perturbation. We find that introducing a distribution of single-ion parameters to mimic \textit{A}-site disorder leads to poorer agreement with the measured spectra, suggesting the local Ni\Sup{2+} electronic structure is largely insensitive to this disorder, within our resolution. Using single-ion values extracted from our model, we calculate the second-order spin-orbit coupling correction to the anisotropic \textit{g}-factor ($g_{\parallel} = 2.26$, $g_{\perp} = 2.27$) and to the effective magnetic moment of Ni\Sup{2+} ($\MuEff=3.2\,\mu_B$). This effective moment is in close agreement to that determined from new magnetic susceptibility measurements ($\MuEff=3.29(4)\,\mu_B$). 
While earlier studies demonstrated spin-liquid behavior in \NCNF{} and establish this material as an important model frustrated magnet, our measurements reveal a remarkably well-preserved NiF\Sub{6} octahedral environment despite the Na\Sup{1+}/Ca\Sup{2+} disorder \cite{krizanNaCaNFrustratedHightemperature2015, plumbContinuumQuantumFluctuations2019}. These new results establish \NCNF{} as a benchmark for single-ion physics on the pyrochlore lattice and demonstrate the ability of RIXS to access important information about low-energy physics, related to magnetism, from measurements of the high-energy single-ion states in strongly correlated materials.

\section{Experiment}
Single crystal samples of \NCNF{} were grown via a modified Bridgman-Stockbarger method in an optical floating zone furnace as described in Ref.~\cite{krizanNaCaNFrustratedHightemperature2015}. A single crystal \NCNF{} of was aligned and cut with the  [111] direction  normal to the sample surface and polished to 0.1\;$\mu$m with diamond paste. Resonant inelastic X-ray scattering (RIXS) and X-Ray absorption (XAS) measurements were carried out on the SIX 2-ID beamline of the National Synchrotron Light Source II \cite{dvorak10MeVResolution2016}. RIXS spectra were collected using a horizontal scattering geometry at a fixed scattering angle $2\theta = 90^{\circ}$ in a specular geometry with linear horizontal ($\pi$) X-ray polarization and the sample [110] direction oriented in the scattering plane. Measurements were performed across the Ni L$_{2,3}$-edge with an energy resolution of of 55\;meV and the F K-edges with an energy resolution of 49\;meV (full width at half max, FWHM). All measurements were taken at room temperature. 

Magnetization measurements were performed using vibrating sample magnetometry (VSM) on a Quantum Design physical property measurement system with an 11.3\;mg sample and a 1000\;Oe field applied parallel to the crystallographic [110] direction.

\section{Results}
\FigMap{s}(a) shows the room temperature Ni L\Sub{3} RIXS intensity in \NCNF{} as a function of incident energy \Ei{} = 850 $\rightarrow$ 860\;eV, and energy loss \eloss{}. There are a series of intense Raman like features between \eloss{} =0 and 4 eV that we identify as \textit{d-d} excitations originating from $\ket{t_{2g}}^m \rightarrow \ket{e_g}^n$ transitions on Ni$^{2+}$.  Integrating the RIXS spectra over the energy transfer yields the partial fluorescence yield (PFY) from which we identify two features: a main peak (MP) at \Ei{}\;=\;853.2 eV and a satellite peak (SP) at \Ei{}\;=\;855.8 eV. \FigMap{s}(b,\,c) show the SP and MP RIXS spectra between \eloss{}\;=\;0\;$\rightarrow$\;4\;eV. The MP (SP) spectrum was fit with a resolution limited Voigt profile, to describe the elastic line, and eight(ten) Gaussians, for the remainder of excitations. The inset in \FigMap{}(b) displays a low intensity peak present in the SP centered around \eloss{} $\approx$ 7\;eV, labeled peak \textit{S} that we will identify below as associated with Ni-F charge transfer. \FigMap{s}(d,\,e) depict the SP and MP RIXS spectra for the Ni L\Sub{2}-edge with \Ei{} = 871.8 and 870.6\;eV respectively, determined by the same method as for the L$_3$-edge spectra. These spectra are fit as well with a resolution limited Voigt profile and nine (eight) Gaussians for the SP (MP). 

\begin{figure}[t!]
  \begin{center}
    \includegraphics{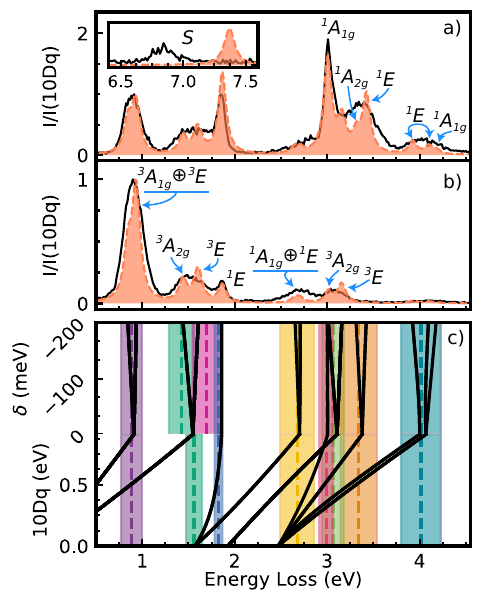}
  \caption{(a,\,b) \NCNF{} SP and MP RIXS scans compared with simulated spectra using single-ion parameters in Table~\ref{tab:LFM_Params} for $D_{3d}$ symmetry and $\lambda_{3d,i}=0$. Inset in (b) shows peak \textit{S} compared with simulated single-ion spectrum. (c) Tanabe-Sugano diagram for 3d$^8$ in $O_h$ symmetry (bottom) and $D_{3d}$ symmetry (top). Vertical dashed lines and shaded regions correspond to the fitted peak positions and FWHM as shown in \FigMap{}\,(b,\,c).}
  \label{fig:LFM_Trigonal}
  \end{center}
\end{figure}
 
Compared to the RIXS spectra of other Ni\Sup{2+} insulators, for example NiO [see Fig.~SI], we observe significantly less fluorescence signal, as well as a near absence of a charge transfer background in the RIXS map of \NCNF{} \cite{Supp}. The comparison  between NiO and \NCNF{} suggest a highly ionic Ni-F bonding nature in \NCNF{}, which in part, results in a larger charge transfer gap between the Ni 3\textit{d} and F 2\textit{p} orbitals. Thus, we begin our analysis of the primarily \textit{d-d} excitations between \elossR{0}{4} in \NCNF{} using a single-ion model.

\subsection{Single-Ion Crystal Field Model}
Our single-ion model for Ni\Sup{2+} in \NCNF{} includes $O_h$ and $D_{3d}$ crystal field potentials, Coulomb interactions parameterized by Slater-Condon integrals (\textit{F}\SupSub{(0,2,4)}{dd}, \textit{F}\SupSub{(0,2)}{dp}, \textit{G}\SupSub{(1,3)}{dp}), and spin-orbit coupling (SOC) of Ni\Sup{2+} 2\textit{p} core ($\lambda_c$) and 3\textit{d} valence orbitals, in the initial and intermediate RIXS states ($\lambda_{3d,i}$, $\lambda_{3d,n}$ respectively). We also include four equivalent, non-interacting Ni\Sup{2+} sites to account for the local axes orientation within the $Fd\bar{3}m$ space-group and crystallographic unit cell. The value of the $O_h$ crystal field splitting parameter 10Dq is fixed by energy of the first energy loss feature in the RIXS map in \FigMap{}(a) \cite{Ghiringhelli}.

The trigonal distortion in \NCNF{} results in a compressed Ni\Sup{2+} octahedral environment along a $C_3$ axis of the $O_h$ point group \cite{krizanNaCaNFrustratedHightemperature2015}. We choose the form of our trigonal potential such that we maintain the orbital quantization axis along a $C_4$ axis of the $O_h$ point group \cite{PryceV, BallhausenLFT}

\begin{align}
    \Pot{D_{3d}}{} = &\; A_2[yz +  xz + xy] \\
   + & \:A_4[yz(r^2-7x^2) + xz(r^2-7y^2) + xy(r^2-7z^2)] \nonumber
\end{align}

where $A_2$ and $A_4$ are the coefficients for the dipolar and quadrupolar terms of the potential \cite{PryceV}. This form of the potential naturally leaves the $\ket{e_g}$ orbitals untouched and splits the $\ket{t_{2g}}$ orbitals into $\ket{a_{1g}}\oplus \ket{e_g^{\pi}}^\pm$ orbitals. 
The matrix elements of this potential in the one electron picture are 
\begin{align}
    &\bra{a_{1g}} \Pot{D_{3d}}{} \ket{a_{1g}} = \sfrac{2}{3} \; \delta \nonumber \\
    &^{\pm}\!\bra{e_g^{\pi}} \Pot{D_{3d}}{} \ket{e_g^{\pi}}\!^{\pm} = -\sfrac{1}{3}\; \delta \\
    &^{\pm}\!\bra{e_g^{\pi}}\Pot{D_{3d}}{} \ket{e_g} = \nu \nonumber 
 \end{align}
where $\delta$ is energy difference between the $\ket{a_{1g}}$ and $\ket{e_g^{\pi}}^\pm$ orbitals, while $\nu$ controls the orbital mixing between the $\ket{e_g^{\pi}}^\pm$ and $\ket{e_g}$ orbitals. For $\nu \neq 0$, the $\ket{e_g^{\pi}}$ orbitals further split into $\ket{e_g^{\pi}}^{+}$ and $\ket{e_g^{\pi}}^{-}$ orbitals. X-ray diffraction measurements suggests that the degree of trigonal compression in \NCNF{} is small \cite{krizanNaCaNFrustratedHightemperature2015}. In addition, the narrow linewidth of the \FrmtSym{1}{E}{}(\FrmtSym{1}{D}{}) peak indicates minimal orbital mixing, as a finite $\nu$ would lead to a splitting of this state. We therefore make use of the approximation which sets $\nu=0$ \cite{PryceV, Supp}. 


\setlength{\tabcolsep}{4.75pt}
\renewcommand{\arraystretch}{1.45}
\begin{table}[b]

\caption{Single-ion parameters of Ni\Sup{2+} in \NCNF{} in $D_{3d}$ symmetry. Slater-Condon parameters are shown as a reduction value from their free-ion values, \textit{F}\SupSub{2}{dd}\,=\,12.234\;eV, \textit{F}\SupSub{4}{dd}\,=\,7.598\;eV, \textit{F}\SupSub{2}{dp}\,=\,7.721\;eV, \textit{G}\SupSub{1}{dp}\,=\,5.787\;eV, and \textit{G}\SupSub{3}{dp}\,=\,3.291\;eV \cite{Ghiringhelli}. The Ni\Sup{2+} spin-orbit coupling value for 2\textit{p} core hole in the intermediate state was fixed to free ion value $\lambda_{c}=11.507$\;eV.} 
\label{tab:LFM_Params}
\begin{tabular}{lllcccccccclll}
 &  &  & \multicolumn{2}{c}{\begin{tabular}[c]{@{}c@{}}Crystal Field \\[-4pt] and SOC\end{tabular}} & \multicolumn{1}{l}{} & \multicolumn{1}{l}{} & \multicolumn{1}{l}{} & \multicolumn{1}{l}{} & \multicolumn{2}{c}{Slater-Condon} &  &  &  \\ \hline \hline
 &  &  & \multicolumn{1}{c||}{10Dq} & 0.91\;eV &  &  &  &  & \multicolumn{1}{c||}{\textit{F}\SupSub{2}{dd}\textit{R}} & 0.78 &  &  &  \\
 &  &  & \multicolumn{1}{c||}{$\delta$} & -200\;meV &  &  &  &  & \multicolumn{1}{c||}{\textit{F}\SupSub{4}{dd}\textit{R}} & 0.77 &  &  &  \\
 &  &  & \multicolumn{1}{c||}{$\lambda_{3d,i}$} & 30\;meV &  &  &  &  & \multicolumn{1}{c||}{\textit{F}\SupSub{2}{pd}\textit{R}} & 0.8 &  &  &  \\
 &  &  & \multicolumn{1}{c||}{$\lambda_{3d,n}$} & 36\;meV &  &  &  &  & \multicolumn{1}{c||}{\textit{G}\SupSub{(1,3)}{pd}\textit{R}} & 0.85 &  &  & 
\end{tabular}
\end{table}

We fit the value of $\delta$ by minimizing the difference between the calculated energy levels for each individual state in $D_{3d}$ symmetry and the fitted peaks in \FigMap{}(b,\,c), with order fixed by the constraint of trigonal compression \cite{PryceV, Supp, delatorreEnhancedHybridizationElectronic2021, discalaElucidatingRoleDimensionality2024}. Importantly, this fit was done with the ground state 3\textit{d} spin-orbit coupling $\lambda_{3d,i} = 0$, which was later used as a free parameter to fit the RIXS spectra. The full fitting procedure is described in the Supplementary Information~\cite{Supp}. \FigTrigonal{}(c) reports the calculated excited state energy levels as a function of the crystal field splitting parameters using fixed values in Table~\ref{tab:LFM_Params}. \FigTrigonal{s}(a,\,b) show the calculated RIXS spectra compared to the SP and MP RIXS spectra, respectively. The missing intensity in the group of states above $\sim$2.5\;eV in the SP spectrum can be attributed to an overlap with fluorescence excitations, which from \FigMap{}(a), peaks at \eloss{}\;$\geq$\;3\;eV. Likewise, the small amount of missing intensity within the \FrmtSym{3}{A}{1g}\;$\oplus$\;\FrmtSym{3}{E}{} peak in the MP spectrum can be attributed to the onset of fluorescence, which pierces through \eloss{}\;=\;10Dq at the incident energy of the MP. The inset in \FigTrigonal{}(a) exhibits a deviation between the data and calculation for the peak \textit{S}. To understand the origin of this peak, we explore the XAS and RIXS spectra at the F K-edge, which provides further insight into the Ni-F bonding nature in \NCNF{}.

\begin{figure}[t!]
  \begin{center}
    \includegraphics{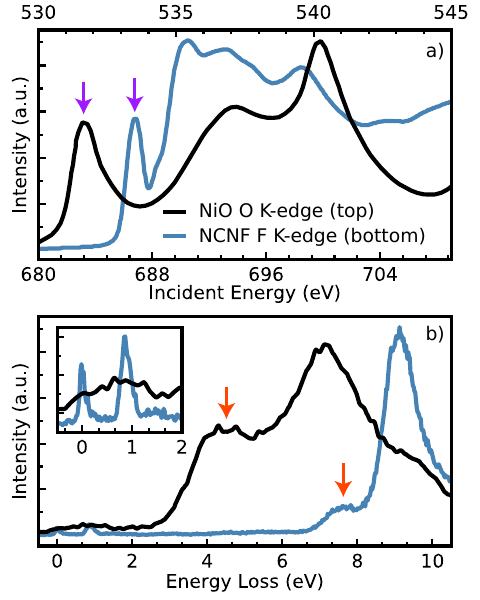}
  \caption{(a)  \NCNF{} Fluorine K-edge XAS (blue, bottom axis) compared to NiO Oxygen K-edge XAS (black, top axis). Purple arrows indicate the excitation energy for RIXS scans in (b). (b) \NCNF{} F K-edge RIXS \& NiO O K-edge RIXS. Orange arrows indicate onset of non-local charge transfer in each material. Inset shows low \eloss{} region with signal of hybridized d-orbitals. NiO data digitized from L.-C. Duda \textit{et al.} \cite{dudaResonantInelasticXRay2006}.}
  \label{fig:XAS_Kedge}
  \end{center}
\end{figure}

\subsection{Evidence of Ni-F Charge Transfer}

\FigKedge{}(a) shows the F K-edge XAS in \NCNF{} which exhibits a distinct pre-edge feature consistent with previous designations of ligand 2\textit{p} - transition metal 3\textit{d} orbital hybridization, paralleling what was seen in the O K-edge XAS in NiO \cite{dudaResonantInelasticXRay2006, SalaEnergySymmetryDd2011}. Signatures of 2\textit{p}-3\textit{d} hybridization are further confirmed by the F K-edge RIXS spectrum, as seen in \FigKedge{}(b), again compared to the O K-edge RIXS spectrum in NiO. Each spectra are dominated by an intense broad peak centered around \eloss{}\;$\approx$\;9\;(7)\;eV, and a high energy shoulder centered at \eloss{}\;=\;7.6\;(4.5)\;eV, indicating the onset of nonlocal charge transfer in \NCNF{} (NiO) \cite{dudaResonantInelasticXRay2006}. The inset of \FigKedge{}(b) shows a low energy peak centered around the 10Dq value reported in Table~\ref{tab:LFM_Params}, consistent with a \textit{d-d} excitation and analogous to that observed in NiO. The greater energy loss of these peaks in \NCNF{} implies a larger charge transfer excitation edge energy, around 7.6\;eV \cite{dudaResonantInelasticXRay2006}. Based on this analysis, we assign peak \textit{S} to the \FrmtSym{1}{A}{1g}(\FrmtSym{1}{S}{}) state and conclude that its energy is lowered due to its proximity to higher energy charge transfer states. To accurately describe this peak necessitates a more complex model, such as an Anderson impurity model, which is beyond the scope of this work.  
\begin{figure}[t!]
  \begin{center}
    \includegraphics{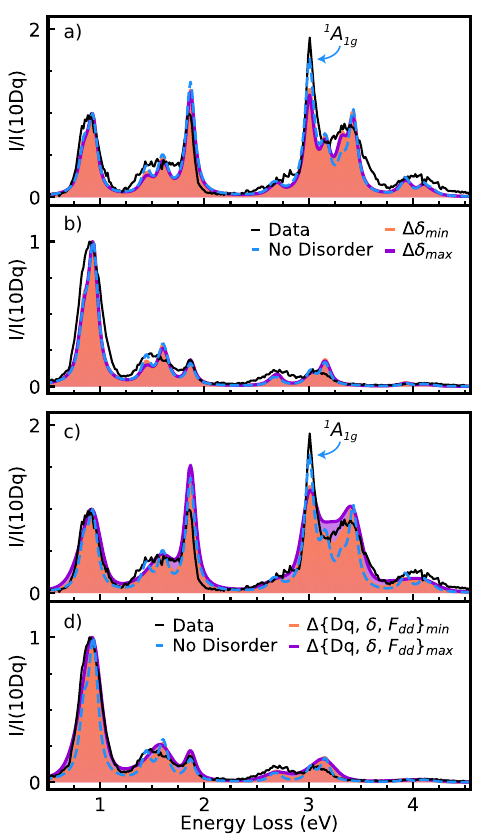}
  \caption{(a,\,b) \NCNF{} SP and MP RIXS scans (black) compared with simulated spectra using single-ion parameters in Table~\ref{tab:LFM_Params} for $D_{3d}$ symmetry (dashed blue), as well as minimal (orange) and maximal (purple) disordered trigonal crystal field where $\Delta\delta_{min}=$\;55\;meV and $\Delta\delta_{max}=$\;150\;meV. (c,\,d) Same as (a,\,b); now the orange and purple spectra include the minimal (55\;mev) and maximal (100\;meV) disordered cubic crystal field ($\Delta Dq$), and Slater-Condon parameters ($\Delta F_{dd}=\Delta\{F_{dd}^{2},F_{dd}^{4}\}$).}
  \label{fig:Disorder}
  \end{center}
\end{figure}
\subsection{Determination of Effective Moment}
Previous measurements of the magnetic susceptibility in \NCNF{} report an effective magnetic moment of $\mu_{ \rm{eff}}=3.7(1)\mu_B$; anomalously large for Ni\Sup{2+} in an octahedral environment with a quenched angular momentum \cite{BallhausenLFT, krizanNaCaNFrustratedHightemperature2015, plumbContinuumQuantumFluctuations2019}. Regardless, the spin-orbit interaction modifies the Land\'e \textit{g}-factor, dependent on the local point group symmetry, thereby influencing the magnetic moment per TM site \cite{PrycePerturbation,BallhausenLFT}. Using the single-ion values extracted from our RIXS model in Table \ref{tab:LFM_Params}, we evaluate the second-order correction to the \textit{g}-factor due to spin-orbit coupling and subsequently calculate the effective magnetic moment in $O_h$ and $D_{3d}$ symmetries.
The general anisotropic Land\'e \textit{g}-factor is then \cite{PrycePerturbation, BallhausenLFT}
\begin{equation}
    g_{ij} = 2(\delta_{ij} - \lambda\Lambda_{ij}) 
    \label{eq:gFactor}
\end{equation}

In $D_{3d}$ symmetry, the \textit{g}-factors are anisotropic and we obtain
\begin{align}
\begin{aligned}
    g_{\parallel} &= 2(1-\lambda\Lambda_{zz}) = 2\bigg(1-\frac{4\lambda}{E_{\Gamma^{b}}}\bigg)  \\
    g_{\perp} & = 2(1-\lambda\Lambda_{xx})= 2\bigg(1- 2\lambda\bigg[\frac{1}{E_{\Gamma^{a}}}+\frac{1}{E_{\Gamma^{b}}}\bigg]\bigg) 
    \end{aligned}
\end{align}

where the \textit{z}-axis of the octahedron is chosen as the high symmetry axis. For \NCNF{} we find $g_{\parallel} = 2.26$ and $g_{\perp} = 2.27$. Then, the effective magnetic moment can be calculated for a general anisotropic \textit{g}-tensor \cite{Carlin}
\begin{align}
    \MuEff & = \mu_B \sqrt{\frac{S(S+1)}{3}(g_{\parallel}^2 + 2g_{\perp}^2)} 
\end{align}
which reduces to the expected expression, ($g\sqrt{S(S+1)}\,\mu_B$), for an isotropic \textit{g}-tensor. In $O_h$ and $D_{3d}$ symmetries, we find an effective magnetic moment per Ni\Sup{2+} of $\mu_{\rm{eff}}=3.2\;\mu_B$ (See Supporting Information for details \cite{Supp}). While $D_{3d}$ symmetry can further modify the effective magnetic moment, the extracted splitting for \NCNF{} ($\delta=-200\;$meV) produces a negligible change to the moment. 

Although the calculated effective moment is still low compared to the reported, we clarify this discrepancy by remeasuring the magnetic susceptibility with a 0.1\;T field applied along the [110] direction. Figure~SIII shows the inverse molar susceptibility $(\chi + \chi_0)^{-1}$, where $\chi_0$ accounts for temperature independent contributions to the susceptibility and was determined by fitting the Curie-Weiss law and minimizing the residuals \cite{mugiranezaTutorialBeginnersGuide2022}. Restricting our Curie-Weiss law fit between 180\,-\,300\;K, where the lower bound is approximately twice $\theta_{CW}$, yields a Curie-Weiss temperature of $\theta_{CW}=-92\;$K and an effective magnetic moment of $\MuEff=3.29(4)\,\mu_B$. The agreement between the extracted effective moment and the value calculated using the second-order perturbative correction further demonstrates the single-ion physics that dominates Ni\Sup{2+} in \NCNF{}.

\subsection{Assessing the Influence of \textit{A}-site Disorder}
Despite the agreement between our model and data in \FigTrigonal{}, the value of $\delta$ in \NCNF{} \textit{may} not be unique. In the rare-earth pyrochlore iridates, the size of the trigonal compression of the TM environment is intricately linked to the ionic radius of the rare-earth ion \cite{witczak-krempaCorrelatedQuantumPhenomena2014, clancyXrayScatteringStudy2016, krajewskaAlmostPureJeff12020, antonovPyrochloreIridatesElectronic2020}. Extending this concept to \NCNF{}, the disordered \textit{A}-site occupation of Na\Sup{1+} and Ca\Sup{2+} would be expected to induce a distribution of trigonal compressions on the Ni\Sup{2+} octahedral environments throughout the crystal. To assess this effect, we average simulated RIXS spectra over a normal distribution of trigonal crystal field splitting \cite{delatorreElectronicGroundState2022}. We constraint the bounds of this distribution using the \FrmtSym{1}{A}{1g}(\FrmtSym{1}{G}{}) state, which given its symmetry, should appear as a resolution limited peak \cite{ruckampOpticalStudyOrbital2005, nagManyBodyPhysicsSingle2020}. In addition, since spin-flip exciations are more likely at the incident energy of the SP, the majority of the intensity for this peak is associated with the \FrmtSym{1}{A}{1g}(\FrmtSym{1}{G}{}) state \cite{ruckampOpticalStudyOrbital2005,haverkortMultipletLigandfieldTheory2012, VeenendaalScatt, nagManyBodyPhysicsSingle2020}. We accordingly set the lower bound for the variation in the trigonal crystal field splitting to the experimental resolution ($\Delta\delta=$\;55\;meV), and the upper bound to the FWHM of the fitted 3\;eV peak ($\Delta\delta=$\;150\;meV). 
\FigDisorder{s}\,(a,\,b) shows these spectra compared to the MP and SP RIXS spectra as well as the calculated spectra with no disorder. 

The spectra with and without a disordered trigonal crystal field qualitatively agree with our data equally well. However, the spectra without disorder results in a better agreement in the intensity of a few peaks, namely the 3\;eV peak. It is clear from the spectra in \FigDisorder{}(b) that the addition of disorder to the trigonal field within our bounds significantly decreases the population of the \FrmtSym{1}{A}{1g}(\FrmtSym{1}{G}{}) state. 

We further assess the influence of the \textit{A}-site disorder on the local Ni\Sup{2+} environment by incorporating a distribution of the cubic crystal field strength ($\Delta Dq$) and Slater-Condon parameters ($\Delta F_{dd}=\Delta\{F_{dd}^{2},F_{dd}^{4}\}$) into the spectra simulated with a disordered trigonal crystal field. The lower bound on the variation of all parameters ($\Delta\{Dq, \delta, F_{dd}\}$) was again set to the experimental resolution. The upper bound however was decreased to 100\;meV such that the calculated 10\textit{Dq} peak fit well within the FWHM of data. \FigDisorder{s}(c,\,d) shows these spectra compared to the MP and SP RIXS spectra as well as the calculated spectra with no disorder. As before, we find no changes in the overall peak positions of the disordered spectra; however, we still observe a reduction in the intensity of the same peaks previously suppressed by the disordered trigonal crystal field alone [see \FigDisorder{s}(a,\,b)], as well as an over simulation of the intensity between 2.5\,-\,3.5\;eV. The overall worse agreement between simulated and observed RIXS spectra in both disorder scenarios suggests that, within the capabilities of our measurement, the disordered \textit{A}-site occupation of Na\Sup{1+} and Ca\Sup{2+} has no significant effect on the local Ni\Sup{2+} electronic environment. 

\section{Conclusion}
In summary, we utilized Ni L-edge and F K-edge RIXS to probe the local electronic structure of Ni\Sup{2+} in \NCNF{}, a unique example of spin-1 Heisenberg antiferromagnet on the pyrochlore lattice. We find the electronic structure of Ni\Sup{2+} in \NCNF{} is well described by a weakly perturbed Ni\Sup{2+} ion in an octahedral crystal field; only a small trigonal compression is needed to fully capture the intensity of some peaks. We extract a nearly isotropic $g$-factor from our single-ion model and find a total magnetic moment of $\mu_{\rm {eff}}=3.2$~$\mu_B$ that is consistent with the paramagnetic moment extracted from a Curie-Weiss fit to the magnetic susceptibility.  We find that including a distribution of single-ion parameters to mimic the Na\Sup{1+} and Ca\Sup{2+} disorder yields poorer agreement to the measured spectra, suggesting that the \textit{A}-site disorder has a minimal impact on the Ni\Sup{2+} electronic structure, within our experimental resolution. Our findings further establish \NCNF{} as a model system for both frustrated magnetism and single-ion physics on the pyrochlore lattice. These results also demonstrate the utility of RIXS to obtain information about low-energy physics (magnetism) from measurements of the high-energy single-ion physics in strongly correlated materials.


\section{Acknowledgments}
Work at Brown University was supported by the National Science Foundation under Grant No. 2429695. This research uses the beamline 2-ID of the National Synchrotron Light Source II, a DOE Office of Science User Facility operated for the DOE Office of Science by Brookhaven National Laboratory under Contract No. DE-SC0012704. 

\bibliography{NCNF.bib}
\pagebreak
\clearpage
\onecolumngrid
\begin{center}
\textbf{\large Supplementary material for ``Stability of the Local Ni\Sup{2+} Electronic Structure to \textit{A}-site Disorder in the Pyrochlore Antiferromagnet \NCNF{}"}
\end{center}

\section{Supplemental Figures}
\renewcommand{\thefigure}{SI}
\begin{figure}[h!]
  \begin{center}
    \includegraphics{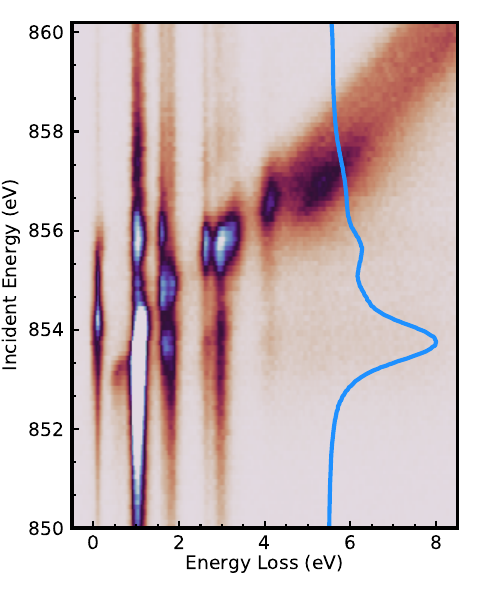}
  \caption{(a) NiO RIXS map across Ni L$_3$ as a function of incident energy $E_i$ and energy loss \eloss{}. The solid blue line shows the XAS as a function of $E_i$.}
  \end{center}
\end{figure}

\renewcommand{\thefigure}{SII}
\begin{figure}[t!]
  \begin{center}
    \includegraphics{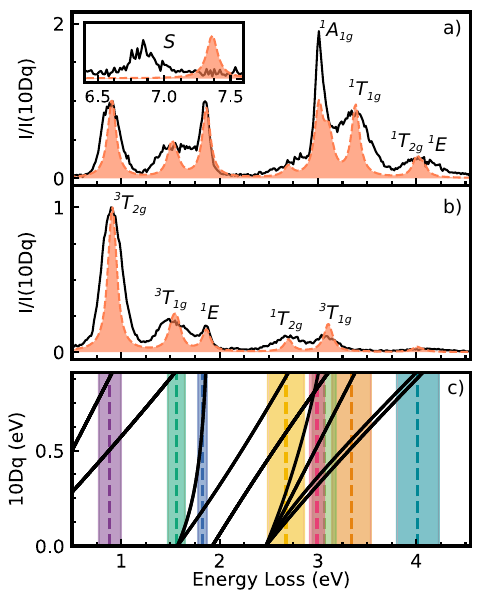}
  \caption{(a)\;\&\;(b) \NCNF{} SP and MP RIXS scans compared with simulated spectra using single-ion parameters in Table~I in the main text for $O_h$ symmetry and $\zeta_{3d,i}=0$. Inset in (b) shows peak \textit{S} compared with simulated single-ion spectrum. (c) Tanabe-Sugano diagram for 3d$^8$ in $O_h$ symmetry. Vertical dashed lines and shaded regions correspond to the fitted peak positions and FWHM as shown in Fig~1(b)\;\&\;(c) in the main text.}
  \label{fig:LFM_Cubic}
  \end{center}
\end{figure}

\renewcommand{\thefigure}{SIII}
\begin{figure}[H]
  \begin{center}
    \includegraphics{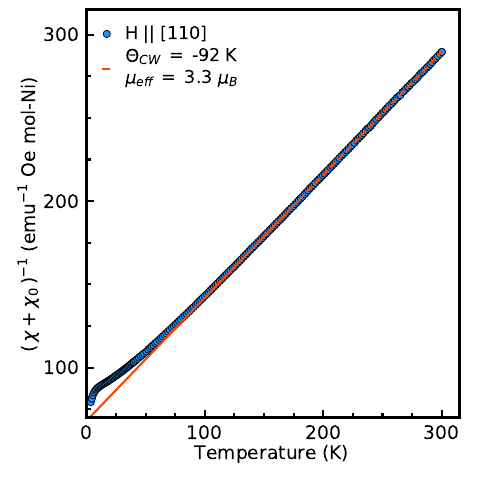}
  \caption{Inverse molar susceptibility $(\chi + \chi_0)^{-1}$ of single crystal \NCNF{} with a 1000\;Oe field applied along the crystal [110] direction. Data plotted every 30 temperature points for clarity. Dashed (Solid) orange line is the Curie-Weiss fit between 180\,-\,300\;K (0\,-\,180\;K).}
  \label{fig:CurieWeiss}
  \end{center}
\end{figure}

\newpage
\section{Electronic Structure Fitting}
\subsection{Ocathedral $O_h$}
To solve for the electronic structure of \NCNF{} in the $O_h$ point group, we consider the peaks fit from the MP and SP, as shown in Fig.~1(b\;\&\;c) in the main text. We note that for the \FrmtSym{3}{T}{1g}(\FrmtSym{3}{F}{}\;) peak, we took the convolution of the two peaks around 1.45\;eV and 1.72\;eV, since two peaks around this region is not possible in $O_h$ symmetry. To obtain the electronic structure parameters shown in Table~I of the main text, we diagonalized the so-called state matrices for each individual state in $O_h$ symmetry which account for the crystal field splitting 10Dq, and Coulomb interactions parameterized by Racah parameters $B = \frac{1}{49}F_{dd}^2-\frac{5}{441}F_{dd}^4$ and $C = \frac{35}{441}F_{dd}^4$, where $F_{dd}^{(2,4)}$ are the intra-orbital Slater-Condon parameters for \textit{d}-electrons. These state matrices are readily available in several texts, but are reproduced here for clarity \cite{GriffithTM, TSII, TSIII}

\renewcommand{\arraystretch}{1.5}
\[
\begin{aligned}
  &
  \begin{array}{|c|c@{\hspace{15pt}}c|}
  \hline
  \FrmtSym{1}{E}{} & t_{2g}^6e_g^2 & t_{2g}^4e_g^4 \\
  \hline
  t_{2g}^6e_g^2 & 8B + 2C & -2\sqrt{3}B \\
  t_{2g}^4e_g^4 & -2\sqrt{3}B & 9B+2C+20Dq \\
  \hline
  \end{array}
  &  \begin{array}{|c|c@{\hspace{15pt}}c|}
  \hline
  \FrmtSym{1}{A}{1g} & t_{2g}^6e_g^2 & t_{2g}^4e_g^4 \\
  \hline
  t_{2g}^6e_g^2 & 16B+4C & \sqrt{6}(2B+C) \\
  t_{2g}^4e_g^4 & \sqrt{6}(2B+C) & 18B+5C+20Dq \\
  \hline
  \end{array}
  \\[0.5em]
  &\begin{array}{|c|c@{\hspace{25pt}}c|}
  \hline
  \FrmtSym{3}{T}{1g} & t_{2g}^4e_g^4 & t_{2g}^5e_g^3 \\
  \hline
  t_{2g}^4e_g^4 & 3B+20Dq & 6B \\
  t_{2g}^5e_g^3 & 6B & 12B+10Dq \\
  \hline
  \end{array}
  & \begin{array}{|c|c@{\hspace{0pt}} c|}
\hline
\FrmtSym{1}{T}{2g} & t_{2g}^4e_g^4 & t_{2g}^5e_g^3 \\
\hline
t_{2g}^4e_g^4 & 9B+2C+20Dq & 2\sqrt{3}B \\
t_{2g}^5e_g^3 & 2\sqrt{3}B & 8B+2C+10Dq \\
\hline
\end{array}
\end{aligned}
\]
\begin{align}
    \FrmtSym{3}{T}{2g} & = 10Dq \nonumber \\ 
    \FrmtSym{1}{T}{1g} & = 12B + 2C + 10Dq \nonumber
\end{align}

We used a differential evolution algorithm to minimize a negative log-likelihood inspired objective function
\begin{equation}
    \label{eq:objFunc}
    \mathcal{L} = \sum_j\bigg[-\log\bigg(\sum_i \exp\bigg\{\frac{(E_{th,i} - E_{obs,j})^2}{2\sigma_j^2}\bigg\}\bigg)\bigg] \nonumber
\end{equation}
where $E_{th}$ and $E_{obs}$ are the calculated and observed energies for a given state, respectively, and $\sigma_j$ is the standard deviation of a fitted peak. The sum over \textit{i} includes the set of calculated energies for a set of states that correspond to a given fitted peak $E_{obs,j}$. For example, the fitted peak around $\sim$\;4\;eV (as shown in Fig.~1\,(b)\,\&\,(c) in the main text) is set to correspond to the calculated energy levels of both the \FrmtSym{1}{E}{}(\FrmtSym{1}{G}{}) and \FrmtSym{1}{T}{2g}(\FrmtSym{1}{G}{}) states. 

\subsection{Trigonal $D_{3d}$}
Following the procedure laid out in the previous section, we use a differential evolution algorithm to minimize the objective function in Eq.~\ref{eq:objFunc} for the following state matrices in $D_{3d}$ symmetry \cite{PryceV}
\[
\begin{array}{|c|c@{\hspace{15pt}}c@{\hspace{15pt}}c|}
  \hline
  \FrmtSym{3}{A}{2g} & \FrmtSym{3}{A}{2g} & \FrmtSym{3}{T}{10\alpha} & \FrmtSym{3}{T}{10\gamma}\\
  \hline
  \FrmtSym{3}{A}{2g} & 0 & 0 & -\sqrt{2}\nu\\
  \FrmtSym{3}{T}{10\alpha} & 0 & 3B+20Dq+\frac{2}{3}\delta & 6B-\sqrt{2}\nu\\
  \FrmtSym{3}{T}{10\gamma} & -\sqrt{2}\nu & 6B-\sqrt{2}\nu & 12B+10Dq+\frac{1}{3}\delta\\
  \hline
  \end{array}
\]
\[
\begin{array}{|c|c@{\hspace{15pt}}c@{\hspace{15pt}}c|}
  \hline
  \FrmtSym{3}{E}{} & \FrmtSym{3}{T}{1\pm\alpha} & \FrmtSym{3}{T}{2\pm} & \FrmtSym{3}{T}{1\pm\gamma}  \\
  \hline
  \FrmtSym{3}{T}{1\pm\alpha} & 3B+20Dq-\frac{1}{3}\delta & \mp\frac{i}{\sqrt{2}}\nu & 6B+\frac{1}{\sqrt{2}}\nu \\
  
  \FrmtSym{3}{T}{2\pm} & \mp\frac{i}{\sqrt{2}}\nu & 10Dq-\frac{1}{6}\delta & \pm\frac{i}{2}\delta \\
  
  \FrmtSym{3}{T}{1\pm\gamma} &6B+\frac{1}{\sqrt{2}}\nu & \mp\frac{i}{2}\delta & 12B+10Dq-\frac{1}{6}\delta \\
  \hline
  \end{array}
\]
\[
\begin{array}{|c|c@{\hspace{15pt}}c@{\hspace{15pt}}c@{\hspace{15pt}}c|}
  \hline
  \FrmtSym{1}{A}{1g} & \FrmtSym{1}{T}{20\alpha} & \FrmtSym{1}{T}{20\gamma} & \FrmtSym{1}{A}{1\alpha} & \FrmtSym{1}{A}{1\beta} \\
  \hline
  \FrmtSym{1}{T}{20\alpha} & 9B+2C+20Dq-\frac{2}{3}\delta & 2\sqrt{3}B-\sqrt{\frac{2}{3}}\nu & -\frac{2}{3}\sqrt{2}\delta & 0\\
  \FrmtSym{1}{T}{20\gamma} & 2\sqrt{3}B-\sqrt{\frac{2}{3}}\nu & 8B+2C+10Dq+\frac{1}{3}\delta & \frac{2}{3}\sqrt{2}\delta & -\sqrt{2}\nu\\
  \FrmtSym{1}{A}{1\alpha} & -\frac{2}{3}\sqrt{2}\delta & \frac{2}{\sqrt{3}}\nu & 10B+5C+20Dq & \sqrt{6}(2B+C)\\
  \FrmtSym{1}{A}{1\beta} & 0 & 0 & \sqrt{6}(2B+C) & 16B+4C \\
  \hline
  \end{array}
\]
\[
\begin{array}{|c|c@{\hspace{15pt}}c@{\hspace{15pt}}c@{\hspace{15pt}}c@{\hspace{15pt}}c|}
  \hline
  \FrmtSym{1}{E}{} & \FrmtSym{1}{E}{\pm\alpha}  & \FrmtSym{1}{E}{\pm\beta}  & \FrmtSym{1}{T}{2\pm\alpha} &\FrmtSym{1}{T}{2\pm\gamma} & \FrmtSym{1}{T}{1\pm} \\
  \hline
  \FrmtSym{1}{E}{\pm\alpha} & 9B+2C+20Dq & -2\sqrt{3}B & \frac{\sqrt{2}}{3}\delta & \frac{2}{\sqrt{3}}\nu & 0 \\
  
  \FrmtSym{1}{E}{\pm\beta} & -2\sqrt{3}B & 8B+2C & 0 & \nu & \mp i\nu\\
  
  \FrmtSym{1}{T}{2\pm\alpha} & \frac{\sqrt{2}}{3}\delta & 0 & 9B+2C+20Dq+\frac{1}{3}\delta & 2\sqrt{3}B+\frac{1}{6}\nu & \mp\sqrt{\frac{3}{2}}i\nu\\
  
  \FrmtSym{1}{T}{2\pm\gamma} & \frac{2}{\sqrt{3}}\nu & \nu & 2\sqrt{3}B+\frac{1}{\sqrt{6}}\nu &8B+2C+10Dq-\frac{1}{6}\delta & \mp\frac{i}{2}\delta\\
  
  \FrmtSym{1}{T}{1\pm} & 0 & \pm i\nu & \sqrt{\frac{3}{2}}i\nu & \pm\frac{i}{2}\delta & 12B+2C+10Dq-\frac{1}{6}\delta\\
  \hline
  \end{array}
\]
\begin{align}
    \FrmtSym{3}{A}{1g} & = 10Dq + \frac{1}{3}\delta \nonumber \\ 
    \FrmtSym{1}{A}{2g} & =12B+2C+10Dq+\frac{1}{3}\delta \nonumber
\end{align}

where $\delta$ is the energy splitting between the $\ket{a_{1g}}$ and $\ket{e_g^{\pi}}$ orbitals, and $\nu$ is the orbital mixing between the $\ket{e_g}$ and $\ket{e_g^{\pi}}$ orbitals. For $\nu \neq 0$, the $\ket{e_g^{\pi}}$ orbitals further split into $\ket{e_g^{\pi}}^{+}$ and $\ket{e_g^{\pi}}^{-}$ orbitals. The subscripts on each state distinguish the orbital occupations in $O_h$ symmetry: $\alpha \equiv \ket{t_{2g}^4e_g^4}$, $\beta \equiv \ket{t_{2g}^6e_g^2}$, and $\gamma \equiv \ket{t_{2g}^5e_g^3}$. For the \FrmtSym{1}{E}{} and \FrmtSym{3}{E}{} matrices, each element should be thought of as a 2\;x\;2 matrix where the only non-zero elements are the overlaps of state with equal parity. For example, 
\[
\bra{\FrmtSym{1}{E}{\pm\alpha}}\Pot{}{}\ket{\FrmtSym{1}{E}{\pm\alpha}} = 
\begin{array}{|c|c@{\hspace{15pt}}c|}
  \hline
  & \FrmtSym{1}{E}{+\alpha} & \FrmtSym{1}{E}{-\alpha} \\
  \hline
  \FrmtSym{1}{E}{+\alpha} & 9B+2C+20Dq & 0 \\
  \FrmtSym{1}{E}{-\alpha} & 0 & 9B+2C+20Dq \\
  \hline
  \end{array}
\]
where \Pot{}{} is the crystal field ($O_h$\;+\;$D_{3d}$) and Coulomb interaction potential. For this fit, in addition to varying the trigonal field splitting, the Slater-Condon parameters were again allowed to vary. 

\section{Effective magnetic moment}
A detailed version of the below discussion can be found in Ref.\,\cite{BallhausenLFT}; a concise summary is reproduced here relevant for \NCNF{}.
To calculate the effective magnetic moment of Ni\Sup{2+} in \NCNF{} we first evaluate the second-order correction to the \textit{g}-factor due to spin-orbit coupling. We consider only the two lowest energy states which largely determine the magnetic properties \cite{BallhausenLFT}. These are the \FrmtSym{3}{A}{2g}(\FrmtSym{3}{F}{}\,) and \FrmtSym{3}{T}{2g}(\FrmtSym{3}{F}{}\,) states. The \FrmtSym{3}{A}{2g} state has components 

\begin{align}
&\FrmtSym{3}{A}{2g} = \begin{cases} 
\ket{\Theta_1} = |\overset{+}{(x^2-y^2)}\overset{+}{(z^2)}| \\
\ket{\Theta_2} = \frac{1}{\sqrt{2}}\big(|\overset{+}{(x^2-y^2)}\overset{-}{(z^2)}| + |\overset{-}{(x^2-y^2)}\overset{+}{(z^2)}|\big) \nonumber \\
\ket{\Theta_3} = |\overset{-}{(x^2-y^2)}\overset{-}{(z^2)}|
\end{cases}
\end{align}

where $|(a)(b)|$ is shorthand for the normalized Slater determinant with $(a)(b)$ along the diagonal, and the $\pm$ overset of each orbital component indicates the spin state. The orbital components of the \FrmtSym{3}{T}{2g} are 

\begin{align}
&\FrmtSym{}{T}{2g} = \begin{cases} 
\ket{\varphi_1} = |(xy)(z^2)| \\
\ket{\varphi_2} = \frac{1}{2}\big(|(xz)(z^2)| + \sqrt{3}|(xz)(x^2-y^2)|\big)  \nonumber \\
\ket{\varphi_2} = \frac{1}{2}\big(|(yz)(z^2)| - \sqrt{3}|(yz)(x^2-y^2)|\big)
\end{cases}
\end{align}
with spin states
\begin{align}
    \psi_1 = \alpha \alpha \quad \quad \psi_0 = \frac{1}{\sqrt{2}}(\alpha\beta + \beta\alpha) \quad\quad \psi_{-1} = \beta\beta \nonumber
\end{align}
where $\alpha$ and $\beta$ denote spin states (+) and (-) respectively. Of the nine possible combinations of $\varphi_i\psi_j$, we only have three states that couple to \FrmtSym{3}{A}{2g} states through the spin-orbit interaction, these are
\begin{align}
    \ket{\Gamma^a} & = \frac{1}{\sqrt{2}}\big(\ket{\Psi_2\psi_{-1}} + i\ket{\Psi_3\psi_1}\big) \nonumber \\ 
    \ket{\Gamma^b} & = \frac{1}{\sqrt{2}}\big(\ket{\Psi_2\psi_0} + i\ket{\Psi_1\psi_1}\big) \label{eq:Gammas} \nonumber \\
    \ket{\Gamma^c} & = \frac{i}{\sqrt{2}}\big(\ket{\Psi_3\psi_0} - \ket{\Psi_1\psi_{-1}}\big) \nonumber
\end{align}
where $\Psi_i$ are linear combinations of $\varphi_i$
\begin{align}
    \ket{\Psi_1} = \ket{\varphi_1} \quad \quad \ket{\Psi_2} = \frac{1}{\sqrt{2}}(\ket{\varphi_2} - i\ket{\varphi_3}) \quad\quad \ket{\Psi_{3}} = \frac{1}{\sqrt{2}}(i\ket{\varphi_2} - \ket{\varphi_3}) \nonumber
\end{align}

Our perturbative potential is the sum of the Zeeman and spin-orbit coupling Hamiltonians, defined as
\begin{equation}
    \hat{\mathcal{H}}^{(1)} = \lambda \textbf{L} \cdot \textbf{S} + \mu_B\textbf{H}\cdot(\textbf{L} + 2\textbf{S}) \nonumber
\end{equation}
where $\mu_B$ is the Bohr magneton and $\lambda$ is the spin-orbit coupling constant defined as $\lambda<0$ for more than half-filled shells \cite{BallhausenLFT}. Then, keeping terms linear in $\lambda$, the second order energy correction is \cite{PrycePerturbation, BallhausenLFT}
\begin{equation}
    E^{(2)} = 2\mu_B(\delta_{ij} - \lambda \Lambda_{ij})S_iH_j \nonumber
\end{equation}
where $\delta_{ij}$ is the Kronecker delta and $\Lambda_{ij}$ is a real, symmetrical, positive definite tensor defines as
\begin{equation}
    \Lambda_{ij} = \sum_{n\neq0}\frac{\bra{\psi_0}\hat{L}_i\ket{\psi_n}\bra{\psi_n}\hat{L}_j\ket{\psi_0}}{E_n-E_0} \nonumber
    \label{eq:Lambda} 
\end{equation}
The indices \textit{ij} go over the \textit{x}, \textit{y}, \textit{z} axes in a given point group symmetry. The general anisotropic Land\`e \textit{g}-factor is then 
\begin{equation}
    g_{ij} = 2(\delta_{ij} - \lambda\Lambda_{ij}) \nonumber
    \label{eq:gFactor}
\end{equation}
In $O_h$ symmetry, the \textit{g}-factors are isotropic and we obtain
\begin{equation}
    g = 2\bigg(1-\frac{4\lambda}{10Dq}\bigg) \nonumber
\end{equation}
In $D_{3d}$ symmetry, the \textit{g}-facotrs are anisotropic and we obtain
\begin{align}
    g_{\parallel} &= 2(1-\lambda\Lambda_{zz}) = 2\bigg(1-\frac{4\lambda}{E_{\Gamma^{b}}}\bigg) \nonumber \\
    g_{\perp} & = 2(1-\lambda\Lambda_{xx})= 2\bigg(1- 2\lambda\bigg[\frac{1}{E_{\Gamma^{a}}}+\frac{1}{E_{\Gamma^{b}}}\bigg]\bigg) \nonumber
\end{align}
Then, the effective magnetic moment can be calculated for a general anisotropic \textit{g}-tensor \cite{Carlin}
\begin{align}
    \MuEff & = \mu_B \sqrt{\frac{S(S+1)}{3}(g_{\parallel}^2 + 2g_{\perp}^2)} \nonumber
\end{align}
which reduces to the expected expression, ($g\sqrt{S(S+1)}\,\mu_B$), for an isotropic \textit{g}-tensor.

\end{document}